\documentclass[10pt, a4paper]{article}
\usepackage[margin=1.25in]{geometry}
\usepackage{float, graphicx}
\usepackage{microtype}
\usepackage[numbers, sort]{natbib}
\usepackage{graphicx}
\usepackage{booktabs}
\usepackage{setspace}
\begin{document}

\title{From Urban Clusters to Megaregions: Mapping Australia's Evolving Urban Regions}

\author{
  Ng, Matthew Kok Ming \\
  \texttt{matthew.ng@unsw.edu.au} \\
  \texttt{City Futures Research Centre}
  \and
  Shabrina, Zahratu \\
  \texttt{zara.shabrina@kcl.ac.uk} \\
  \texttt{King's College London}
  \and
  Sarkar, Somwrita \\
  \texttt{somwrita.sarkar@sydney.edu.au} \\
  \texttt{University of Sydney}
  \and
  Han, Hoon \\
  \texttt{h.han@unsw.edu.au} \\
  \texttt{City Futures Research Centre}
  \and
  Pettit, Christopher \\
  \texttt{c.pettit@unsw.edu.au} \\
  \texttt{City Futures Research Centre}
}

\maketitle

\begin{abstract}
This study employs percolation theory to investigate the hierarchical organisation of Australian urban centres through the connectivity of their road networks. The analysis demonstrates how discrete urban clusters have developed into integrated regional entities, delineating the pivotal distance thresholds that regulate these urban transitions. The study reveals the interconnections between disparate urban clusters, shaped by their functional differentiation and historical development. Furthermore, the study identifies a dichotomy of urban agglomeration forces and a persistent spatial disconnection between Australia's wider urban landscape. This highlights the interplay between urban densification and peripheral growth. It suggests the need for new thinking on potential integrated governance structures that bridge urban development with broader social and economic policies across regional and national scales. Additionally, the study emphasises the growing importance of national coordination in Australian urban development planning to ensure regional consistency, equity, and productivity.
\end{abstract}   

\section{Introduction}
Cities are complex systems characterised by the dynamic process of merging and partitioning of regions and communities \citep{arcaute2016cities}. They are continuously shaped by geographical, political, and historical factors, creating diverse urban environments. The expansion of cities is supported by infrastructure networks, which reflect the spatial organisation and functional differentiation that characterise urban areas \cite{benguigui2007dynamic, masucci2013limited, zhao2017spatial}. The road network plays a pivotal role in this differentiation, serving as a vital reflection of urban functions and a primary driver of city expansion. Its structures are directly shaped by historical patterns and usage, offering valuable insights into future growth trajectories \cite{batty2009cities, arcaute2015constructing}.

Indeed, \cite{zhao2017spatial} highlighted a direct correlation between network expansion and growth trajectories that follow Gibrat’s Law \cite{eeckhout2004gibrat}, a pattern previously confirmed by \cite{luckstead2014world} in their comparison of city conformations to established growth frameworks (e.g., Zipf's Law). Similarly, \cite{masucci2013limited} demonstrated that key topological features, such as network growth and average nodal connectivity, can be used to provide precise delineations of urban areas. This finding is supported by \cite{benguigui2007dynamic} in their models of urban partitioning.

However, while these studies provide useful insights into quantifying growth patterns, they leave the structured relationships and hierarchical order between urban clusters less explored, which represents a significant area for future research.

The application of percolation theory provides a valuable means of investigating the connectivity and hierarchical organisation of cities, offering a link between the insights gained in the aforementioned research \cite{stauffer1979scaling, makse1998modeling}. Originally a concept in physics, it has found useful applications in urban studies. By incrementally connecting network elements, it is possible to identify the hierarchical relationships that exist within urban systems. This ultimately enables the characterisation and regionalisation of urban centres at different scales. The process thus organises urban network connections, including urban centres, cities, and towns, into a structured hierarchy based on the size of each cluster within the network. In this framework, each segment or cluster represents an urban cluster and the corresponding regional systems, collectively reflecting the natural organisation of the urban network.

The potential of this approach was demonstrated by \cite{arcaute2015constructing}, who used percolation to map the hierarchical organisation of towns and cities in Britain. They identified critical network connection points that reflected well-known existing hierarchies in the United Kingdom, revealing patterns and clusters that have been influenced by both natural and historical socioeconomic divisions. It is of particular significance that their work illuminated the manner in which these divisions (or percolation transitions), which may be defined as the points at which disparate network segments converge to form discernible centres, can be attributed to the expansion of the country's urban centres and their associated functions. These transitions were deemed crucial for understanding urban hierarchies and regional divisions, further informing development planning and socioeconomic policy. 

Building upon this, \cite{raimbault2019multi} applied percolation analysis, at a more regional scale, to population and transport networks across Western Europe. Their work shed light on the intricate interconnectivity of European urban regions, delineating expansive mega-urban areas and identifying a multitude of diverse activity hubs as their fundamental constituents. Significantly, the study also uncovered distinct transition states for different interacting network types, relating to the different economic and population hubs across Europe. The research proposed a novel framework for unravelling the interactive relationships between urban function, population distribution, and movement within urban systems that were previously considered distinct.

\subsection{The evolution of Australian urban centres}
Australia's geography and diverse urban centres offer a unique case study for applying percolation to better understand the relationships and functional connections between its numerous urban centres. Historical records indicate that Australia's urban development has been shaped by three key factors: agriculture, mining, and the growth of transportation infrastructure \cite{freestone2003functions}. In particular, the rapid expansion of railway networks during the Industrial Revolution has had a significant impact on the country's urban landscape. By 1910, most of Australia's major urban centres were already established, primarily located along the coast or near rail networks \cite{BITRE_2014}. However, the pattern of urban growth in Australia has since shifted significantly. In recent times, there has been a decline in the relative population of regional towns and cities, while the population in capital cities has grown and become denser, with car travel becoming prevalent in these urban areas \cite{troy2004structure, coffee2016visualising, wilkinson2022federalism}.

Furthermore, there have been notable shifts in the functional roles of urban centres throughout the country\cite{freestone2003functions}. In particular, research by \cite{freestone2003functions} on new urban hierarchies provides an economic and functional classification of Australia's contemporary urban centres (ref. Table \ref{tab:freestone}). This analysis identifies distinct profiles among these urban areas. The presented framework allows for the identification of the distinctive characteristics and roles of diverse urban regions. Additionally, the analysis emphasises the evolving roles and diversification of towns, underscoring the changing relationship among Australia's urban centres and their centralising effect in response to modern economic changes.

\begin{table*}
\small
    \centering
\phantom{~}\noindent
\begin{tabular}{p{0.1\textwidth}p{0.25\textwidth}p{0.55\textwidth}}
\hline
Cluster & Archetype & Description and Example \\
\hline
1     & Administration and Defence & Centres with high public sector employment, important for government and defence. Examples: Canberra, Darwin, Townsville, Crib Point \\
2     & Power Generation & Centres specialising in utility services, often near resource-rich areas. Examples: Hunter Valley, Gippsland, Biloela, Collie \\
3     & Diversified & Major urban centres with varied economies. Examples: All mainland state capitals \\
4     & Regional Centres & Centres with community services and agriculture, serving large regions. Examples: Armidale, Dubbo, Goulburn, Taree, Wagga Wagga \\
5     & Service & Smaller centres focused on service and agriculture. Examples: Southwestern Victoria to Queensland \\
6     & Rural Processing and Production & Manufacturing centres tied to rural industries. Examples: Rural processing and production centres \\
7     & Tourism and Leisure & Centres with growth in recreation and tourism. Examples: Coastal centres in NSW; Australian alpine region (Bright, Perisher Valley, Jindabyne, Thredbo) \\
8     & Agricultural Service & Centres with agricultural and manufacturing employment. Examples: Irrigation towns, food processing centres \\
9     & Mining Service & Mining-centric towns with additional roles. Examples: Broken Hill, Kalgoorlie \\
10    & Transportation & Centres focused on transportation and tourism. Examples: Queensland transportation centres, Great Barrier Reef service towns \\
11    & Mining & Towns almost entirely dependent on mining. Examples: Coal centres in Queensland, 'goldfields' towns, Weipa, Alyangula, Leigh Creek, Roxby Downs, Rosebery, Zeehan \\
12    & Aboriginal Land Trust & Economies dominated by the public sector on Aboriginal land. Examples: Towns north of the Tropic of Capricorn \\
13    & Aboriginal Remote & Communities with high unemployment, focused on community services. Examples: Cherbourg Aboriginal Community, remote indigenous reservations \\
\hline
\end{tabular}
    \caption{Functional archetypes of Australian urban centres as adapted from \cite{freestone2003functions}.}
    \label{tab:freestone}
    \normalsize
\end{table*}

These changing spatial patterns, driven by economic and infrastructural shifts, have been demonstrated to have tangible effects on development and population dispersion. As illustrated by \cite{slavko2020city}, the influence of these urban configurations on the spatial distribution of population patterns across Australian capital cities, aligning with \cite{raimbault2019multi}'s findings on the relationship between urban infrastructure and social dynamics. \cite{sarkar2020measuring} quantified these functional relationships across Australian suburbs using centrality measures, thereby demonstrating that the polycentricity of functional centres and population cores is primarily driven by transit-led infrastructure development. While these studies contribute to our understanding of urban growth patterns and centralities, there is still a need for a comprehensive description of the relationships and hierarchical order between Australia's distinct urban clusters.

This study is an exploratory examination of Australia's urban hierarchical structures, as revealed by its road network. Our research aims to extend the knowledge gained from previous studies by exploring the connectivity of the road network in uncovering the network partitioning of different urban centres throughout the country. Specifically, we focus on revealed urban clusters, which are defined through their nodal connectivity. We explore these structures through the lens of percolation theory, seeking to move beyond traditional statistical methods typically used in Australia. Our approach aims to uncover more nuanced relationships between the country's many different urban centres and highlight their existing hierarchical structures. The findings of this research are expected to offer new insights into the country's urban landscapes, which may have utility in substantiating urban growth trends and developing sustainable urban planning policies in future.

\section{Methodology}
\subsection{Data}
\begin{figure}[h]
    \centering
    \includegraphics[width = 1\textwidth]{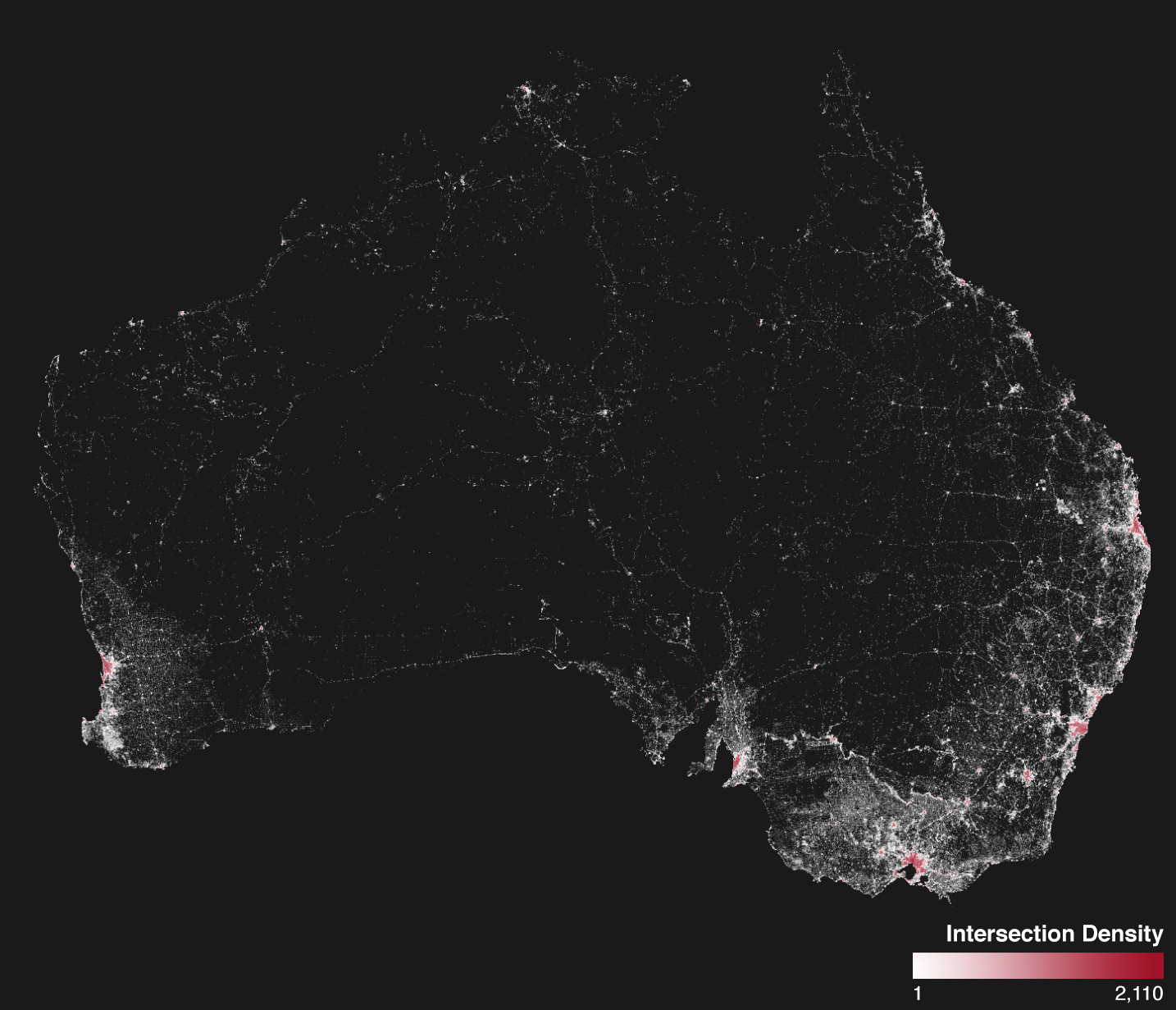}
    \caption{Intersection density of the Australia road network binned in 150m grids}
    \label{fig:int_density}
\end{figure}

The analysis employed the road network of Australia, sourced from OpenStreetMap (OSM). OSM is an open-source project that provides urban datasets, created through crowdsourcing, to the general public. Figure \ref{fig:int_density} visualises the spatial distribution and intersection density of the dataset, which has been rasterised at 150-metre grids. In the processing stage, the data was simplified to reduce the number of vertices and to remove nodes that do not convey any topological meaning. This is in accordance with the specifications set by \cite{arcaute2016cities}. The initial dataset, comprising over 9 million points, was reduced to approximately 1.9 million intersection points, each representing an individual node within the network graph. Moreover, to guarantee network connectivity, an iterative nodal intersection function was devised and implemented on the processed dataset to guarantee network connectivity. This ensured that the degree of network connectivity equalled 1, indicating that all nodes in the network are connected to each another. Each node was subsequently indexed, and its adjacency to other connected nodes was defined through edge weightings corresponding to network distances within a distance matrix. This constructed network served as the basis for the subsequent percolation analysis.

\subsection{Network Percolation}
Percolation theory examines the additive linking of network intersections as distance thresholds increase, with each threshold marking a transition point where previously unconnected segments merge into larger, connected clusters \cite{stauffer1979scaling, makse1998modeling}. At any given threshold, the largest connected units, constrained by the distance threshold $d_0$, are aggregated and considered as single urban clusters. These clusters represent the maximal network-continuous links for that threshold \cite{arcaute2016cities, fluschnik2016size, huynh2019continuum}.

In this study, we conducted a percolation analysis on interconnected intersection points using a distance-based clustering algorithm, following the methods described by \cite{arcaute2015constructing}. Clusters are initiated at random nodes and subsequently expanded through the recursive addition of new links. The formation of a cluster is contingent upon the proximity of two nodes---denoted by $v_k$ and $v_i$---to one another, with the requisite threshold being $d_0$. This is represented by the condition $d(v_k, v_i) < r_0$. In this analysis, the percolation radius, denoted by $r_0$, was set within a range of 100 metres to 30,000 metres, with distance increments of 20 metres. Given the expansive extent of the study area, the formation of a single cluster was not reached, even at the maximum threshold of 30,000 metres. Consequently, several isolated clusters remained at the maximal threshold set for the analysis.

\section{Results}
\begin{figure}[ht]
    \centering
    \includegraphics[width= 1\textwidth]{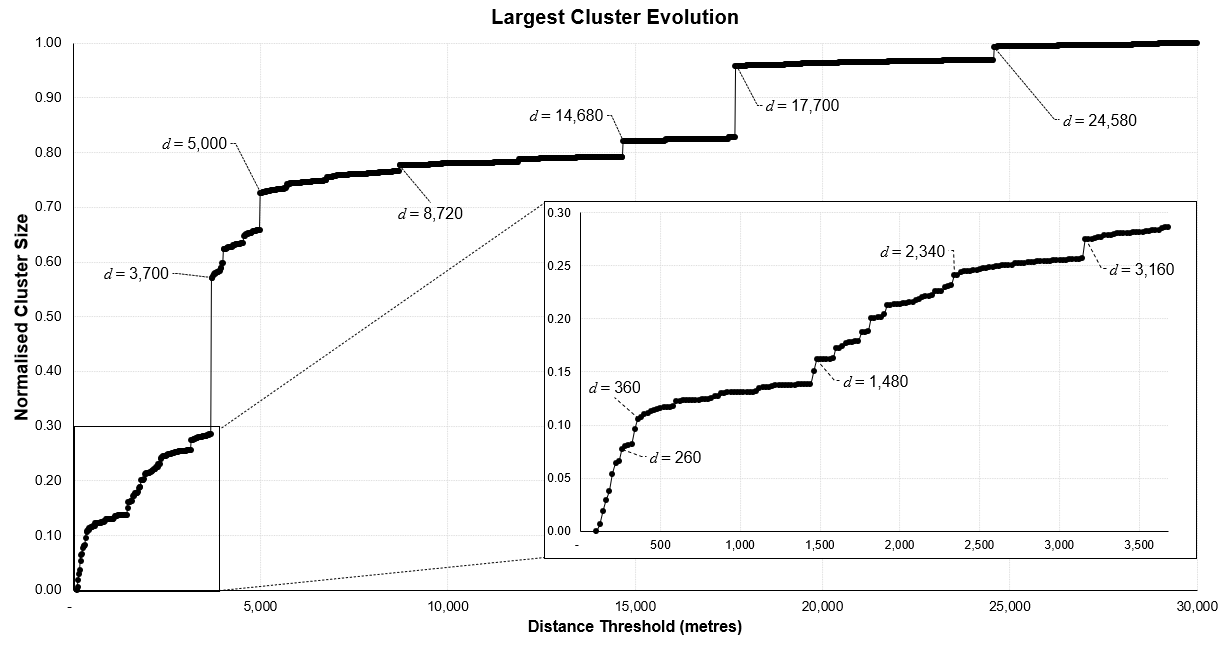}
    \caption{Transition state of the largest cluster's size across all percolation distance thresholds}
    \label{fig:1}
\end{figure}

We track the evolution of the largest cluster across varying distance thresholds, revealing a series of percolation transitions that lead to the formation of distinct urban clusters nationwide. The progression of the largest cluster, shown in Figure \ref{fig:1}, demonstrates how regions merge at different percolation radii $r_0$, indicated by the number of nodes. This analysis quantifies the transformation of urban clusters, highlighting the correlation between the size of the largest cluster and its expansion throughout the system.

\begin{figure}[ht]
    \centering
    \includegraphics[width = 1\textwidth]{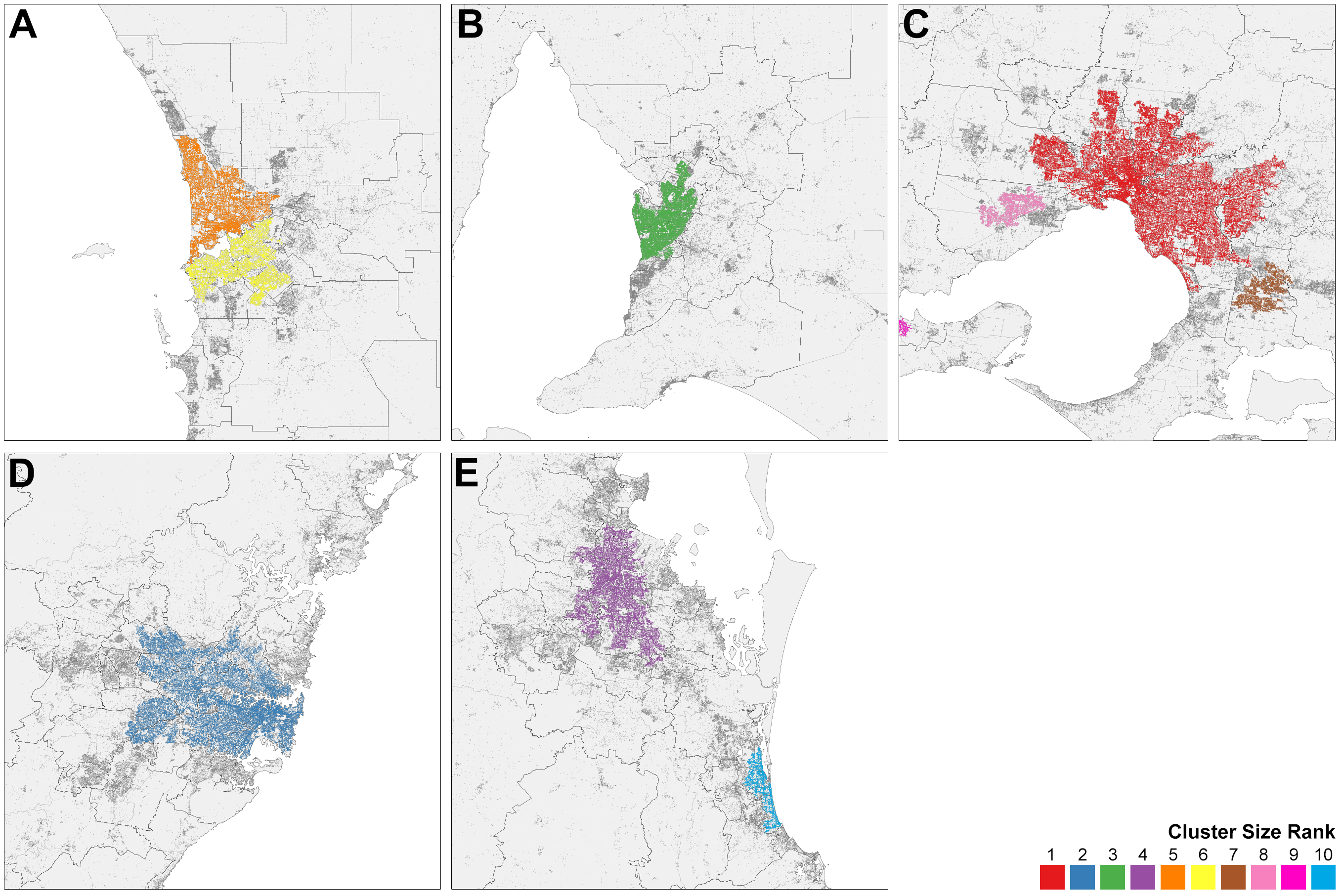}
    \caption{The largest clusters at the first transition phase relates to the country's largest urban areas}
    \label{fig:2}
\end{figure}

The initial transition phase, occurring at distance of 260m, unveils most capital cities (ref. to Figure \ref{fig:2}). The largest clusters are identified for Melbourne, Sydney, and Adelaide (ref. Figure \ref{fig:2}A - Figure \ref{fig:2}D). \cite{arcaute2015constructing} refers to this threshold as the '\textit{system of cities}', which aligns with \cite{freestone2003functions}'s archetype of '\textit{diversified urban centres}'. These centres are characterised by their denser populations, dense urban structures, and varied economies. In the state of Victoria, it is evident that distinct, sizeable clusters are present on the outskirts of Melbourne, encompassing areas such as Werribee, Cranbourne, and Geelong. The distinction between the city centre and these suburbs, despite their relative proximity, is highlighted by this separation. These suburbs may be regarded as significant cores in their own right. A notable separation between the northern and southern parts of Perth is also observed (ref. Figure \ref{fig:2}A), with the city topographically divided by Swan River. It is interesting to note that this separation is not as pronounced in Sydney, which has a more interconnected urban structure despite being separated by Sydney Harbour. In Queensland, the analysis reveals that Brisbane and the Gold Coast are the largest urban centres in the state (ref. Figure \ref{fig:2}E). It is worth noting that the core of Brisbane, particularly in comparison to other capital city clusters at this distance threshold, appears to be considerably smaller than expected, especially when considering the extent of the statistical boundary of Greater Brisbane. Similarly, Adelaide presents a clearly defined boundary to its city extent, which is topologically constrained by the expanse of national parks to its south.

At a distance threshold of 360m, network percolation shows the expansion of most capital city clusters, accompanied by the emergence of several smaller secondary cities. In Victoria, Melbourne's city centre undergoes significant expansion, effectively integrating most of the suburbs to the south-east and north of Melbourne's city centre. In Victoria, Melbourne's city centre undergoes significant expansion, effectively integrating most of the suburbs to the south-east and north of Melbourne's city centre. In addition, centres such as Wyndham Vale and Truganina merge to form the largest cluster in Victoria. At this distance, the secondary cities Ballarat and Geelong emerge as the other major agglomerations in Victoria. This highlights their clear separation from Melbourne city. It is worth noting that these clusters were historically part of the Victorian Gold Cities, which were established as the state's major manufacturing centres in part due to their connection to the railway \cite{freestone2003functions}. In New South Wales, however, the main city of Sydney has not expanded significantly. Instead, the secondary city of Newcastle emerges as the second most prominent cluster. In this transitional phase, Canberra also begins to form a large, distinct cluster, indicating its urban importance in the region's network configuration. In Western Australia, North and South Perth merge. No other significant clusters are identified in Queensland and South Australia at this distance threshold, suggesting a relative separation of other peripheral urban centres within the region.

\begin{figure}[H]
    \centering
    \frame{\includegraphics[width = 1\textwidth]{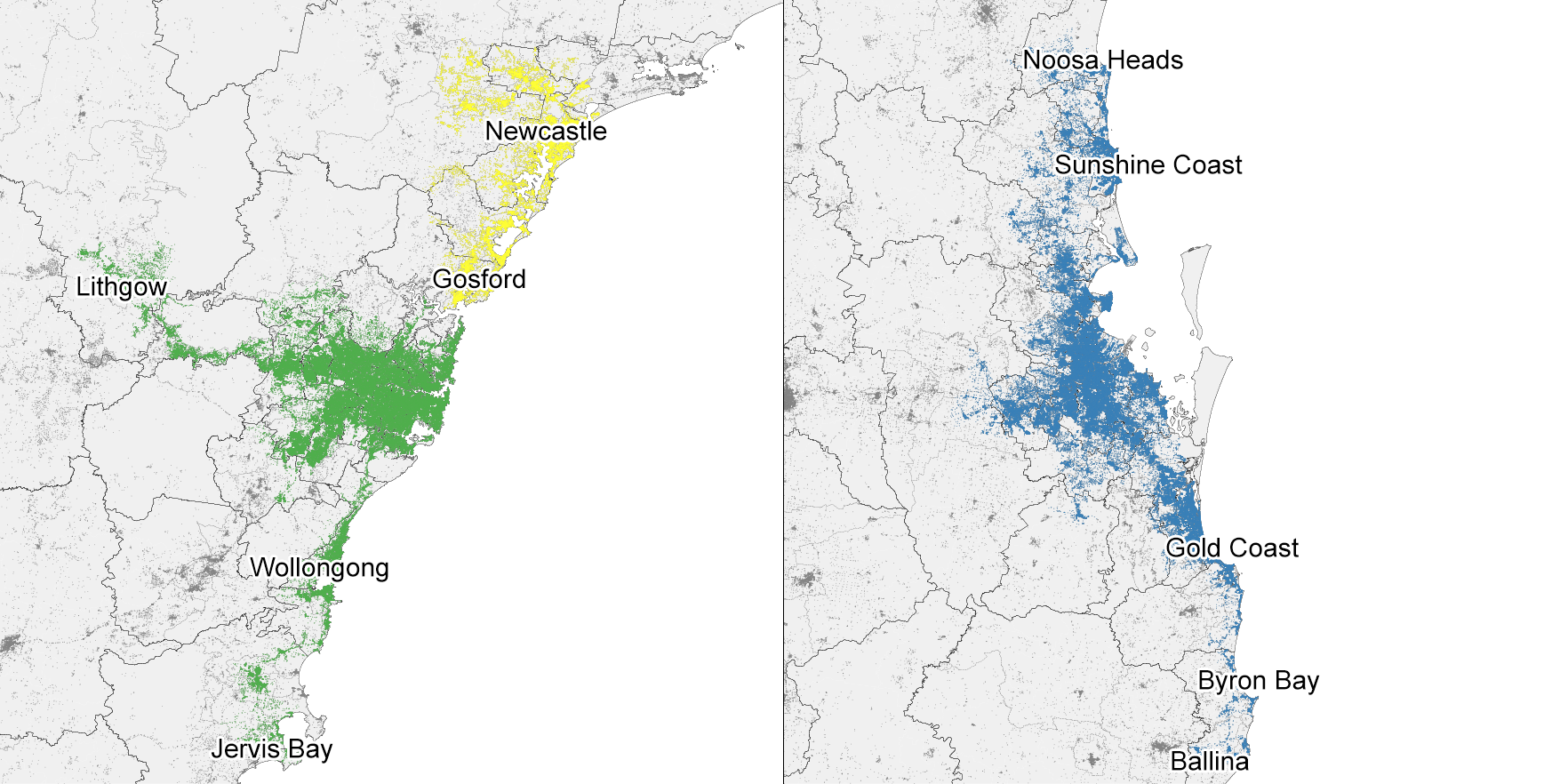}}
    \caption{Cluster differences along Australia's eastern coast at $d = 1,480m$.}
    \label{fig:3}
\end{figure}

At the 1,480m threshold, the analysis reveals the country's many \textit{regional systems} through the agglomeration of the main coastal cities. In Victoria, a notable urban cluster emerges, extending from Bendigo to Castlemaine and merging with the Melbourne metropolitan area. Ballarat, however, remains a separate cluster, distinct from the Melbourne metropolitan area. In Queensland, Townsville emerges as a new significant cluster; in the Northern Territory, Darwin city becomes visible as a distinct cluster at this threshold, marking its first appearance in the transition phases. Both Townsville and Darwin, originally seaports, have diversified their economies to include defence facilities and tourism, and both now stand as distinct clusters of comparable size to other Australian cities. An extensive cluster extending from Noosa Head to Ballina in the city of Brisbane is also observed as the second largest cluster in the system. In New South Wales, despite their geographical proximity, there remains a persistent disconnect between the Sydney-Shoalhaven cluster and the Gosford-Newcastle cluster. It is notable that these two clusters in NSW form part of the state's Greater City Commissions "six cities" regional plan \cite{GCC_2022-key}. These areas are classified by \cite{BITRE_2014} as 'coastal commuter' areas, located in transition zones between regional and urban areas. However, their separation suggests complex patterns of urban growth that may be influenced by physical barriers, differences in economic activity or differences in urban planning and development policies. In Western Australia and South Australia, the main clusters in Perth and Adelaide continue to grow.

At the 2,340m distance threshold, the analysis identifies the emergence of a distinct group of satellite towns, each with a population size between 500 and 1,500 people --- a finding similar to that shown by \cite{BITRE_2014}. This threshold highlights the geographic distribution and connectivity of smaller communities in relation to larger urban centres. In Victoria, the Melbourne cluster begins to extend beyond state boundaries, while smaller, distinct clusters around Marong, Mount Gambier, Hamilton and Portland are observed as separate entities, reflecting the diverse urban landscape within the state. Along the east coast, spanning New South Wales and Queensland, a significant urban cluster extends from Maryborough to Shoalhaven, reflecting the dense urbanisation along Australia's east coast, with Beerburrum and Tumbulgum emerging as distinct clusters from this amalgamation. In contrast, Western Australia and South Australia do not show significant expansion of primary capital clusters. This finding is indicative of Perth's central role within its state and Adelaide's relative isolation from other major urban centres. No significant clusters are found in the Northern Territory at this threshold.

\begin{figure}[H]
    \centering
    \includegraphics[width = 0.8\textwidth]{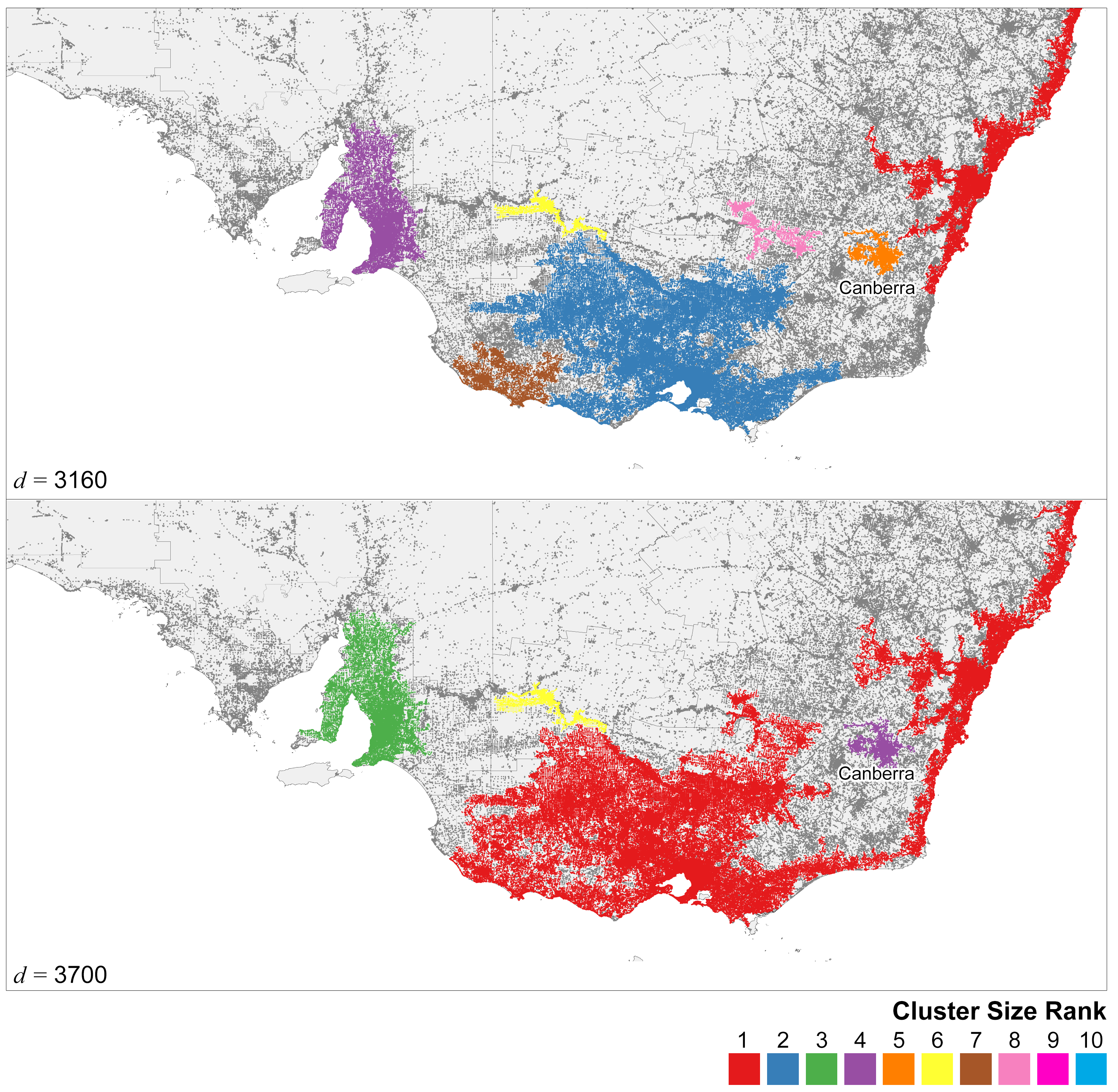}
    \caption{Cluster formation between $d = 3,160m$ and $d = 3,700m$, highlighting the separation of Canberra from NSW and Victoria.}
    \label{fig:east3160}
\end{figure}

Between the distance thresholds of 3,160m and 3,700m (see figure \ref{fig:east3160}), a notable transition occurs along the south-east coast of Australia. During this phase, the main clusters in New South Wales and Victoria merge to form an extensive northeastern crescent covering the area from Brisbane to Melbourne. More than 80 per cent of Australians live in this cluster. This clustering represents a significant shift in the spatial dynamics of these regions. In particular, wve see the merging of cities within the country's notable \textit{"wheat-sheep belt"} into its major urban clusters, potentially delineating the country's modern economic geography. It should be noted, however, that despite the large distance thresholds, distinct clusters continue to exist independently. In particular, Canberra remains a distinct cluster, further emphasising its separate geographic and infrastructural identity. Despite its proximity to both Sydney and Melbourne, Canberra remains distinctly separate from this integrated urban system. Canberra's distinct geographic and infrastructural characteristics prevent it from being integrated into the Sydney-Melbourne supercluster, emphasising its independent spatial identity. Similarly, Adelaide maintains its separation from the Melbourne and Sydney superclusters. Both of these separations are particularly striking given their close physical distance from their nearest metropolitan areas. However, their separation reflects unique factors in their regional positioning.

\begin{figure}[ht]
    \centering
    \includegraphics[width = 1\textwidth]{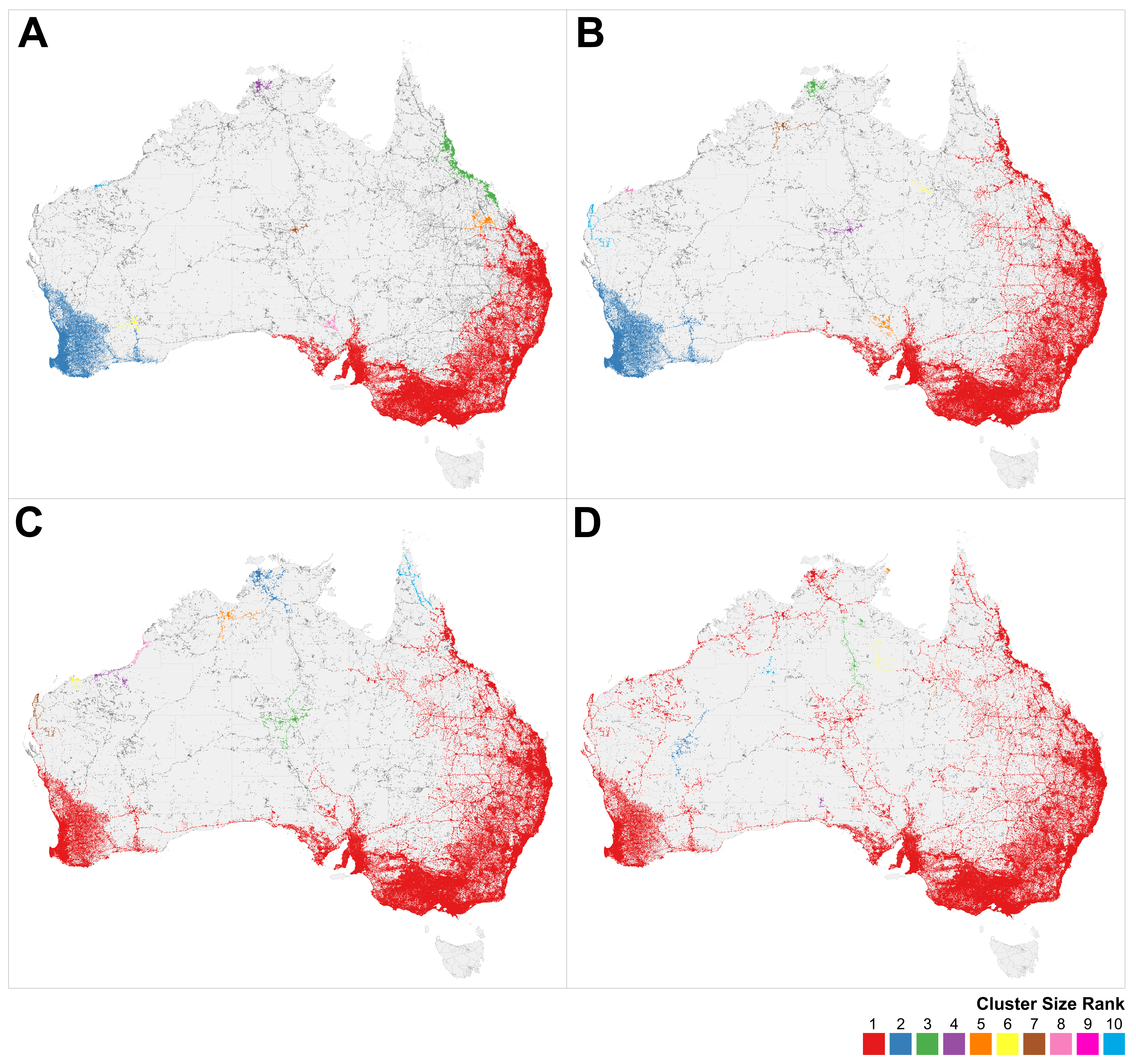}
    \caption{Cluster changes at distances 8,720m (A), 14,680m (B), 17,700m (C), and 24,580m (D).}
    \label{fig:t8}
\end{figure}

At this threshold, the emergence of smaller, more isolated cities within the system becomes apparent. For example, smaller cities such as Townsville and Cairns in Queensland remain separate, further illustrating the diverse spatial dynamics of Australia's urban landscape. Centres such as Alice Springs, Kununurra, Karratha, Broome and Port Augusta also stand out as distinct clusters. These smaller cities, identified by Freestone as historic agricultural and mining service towns and transport hubs, reflect the post-industrial development of these regions. They continue to play a vital role in supporting regional economies and connecting remote areas, despite their isolation from the larger urban clusters along the south-east coast.

The final stages of network transitions in our study are observed between distances of 8,720m and 24,580m. At a distance of 8,720m, the analysis reveals a clear separation between eastern and western Australia, illustrating the notable east-west divide in the country. Smaller clusters in the Northern Territory and North Queensland are observed, relating to a separate amalgamation that encompasses the region from Cairns to Mackay (ref. Figure \ref{fig:t8}A). At a distance of 14,680m (ref. Figure \ref{fig:t8}B), these clusters coalesce to form the largest cluster in eastern Australia, still remaining distinct from those in Western Australia and Northern Australia. In this phase, towns in central Australia, including Alice Springs, begin to emerge. As the transition progresses past the 17,700m threshold, most of Australia's urban centres amalgamate into a single extensive cluster, with remaining separation of smaller towns across the country, such as Karratha, Broome, and Mount Isa (ref. Figure \ref{fig:t8}C). It is interesting to note these final transition phases also uncover the remote Aboriginal Land Trust areas, in addition to historic mining centres. 

\begin{figure}[ht]
    \centering
    \includegraphics[width=1\textwidth]{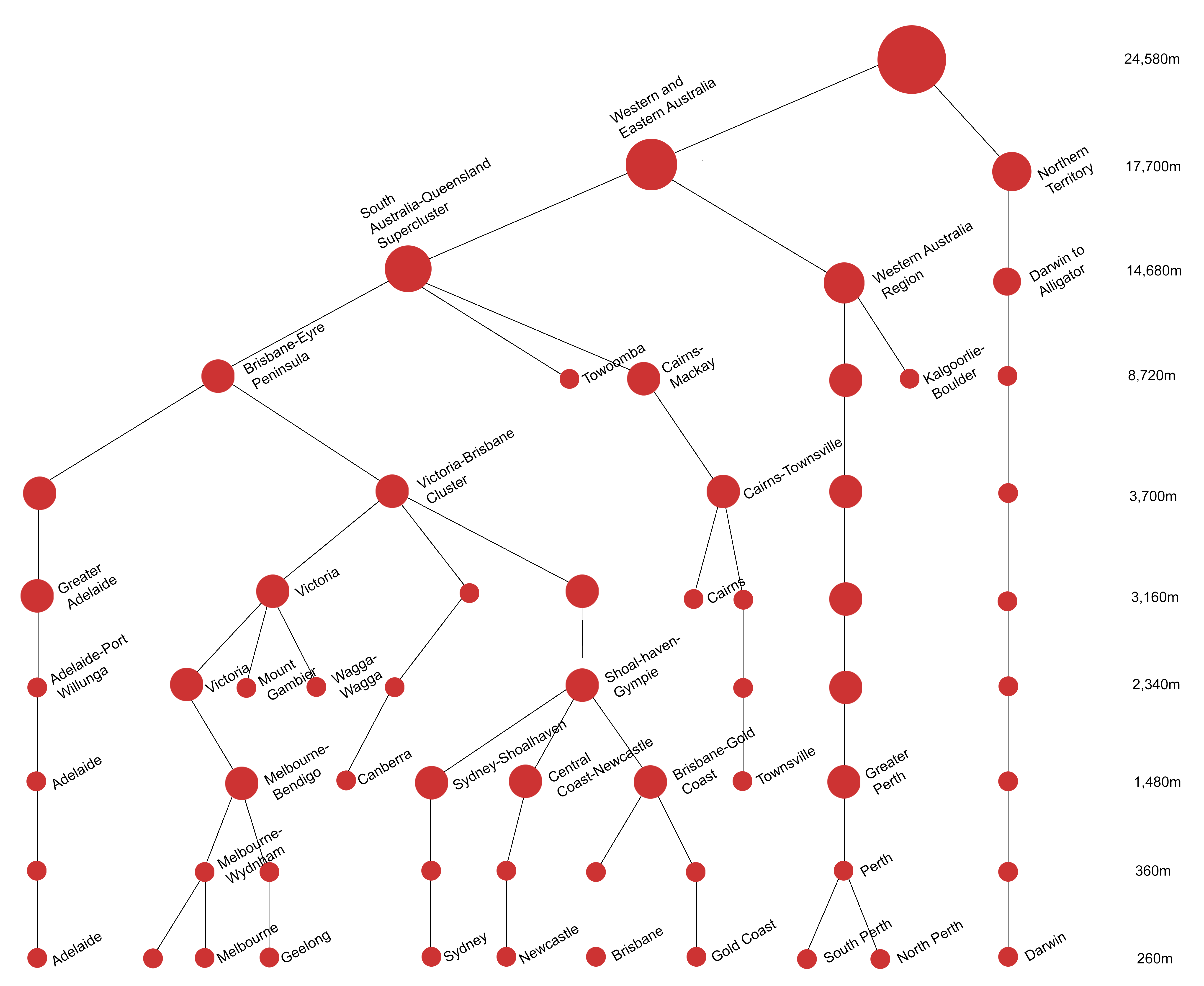}
    \caption{A simplified illustration of how clusters change and merge across each distance threshold, highlighting the progression from localised groupings to broader regional clusters.}
    \label{fig:t9}
\end{figure}

In the final transition phase, the majority of Australia converges into largely a single cluster. 
A simplified visualisation of the largest cluster, illustrating how the final percolation transition relates to broader regional groupings and constituent cities, is shown in Figure \ref{fig:t9}. No further significant transition phases were identified as the percolation analysis concluded at a distance of 30,000m. A reduced visualisation of the largest cluster is shown in Figure \ref{fig:t9}.

\section{Discussions}

Australia's population is expected to increase by around 11.3 million by 2056, with 75 per cent of this growth concentrated in the major cities of Sydney, Melbourne, Brisbane and Perth. This demographic evolution is likely to reinforce the current urban development trajectory, with people settling particularly along the country's east coast. Cities and towns in this area appear to be evolving into a complex network of suburban and urban centres, potentially leading to the formation of a crescent-shaped mega-metro regional hybrid along the east coast, from Melbourne-Geelong to Sydney-Newcastle and south-east Queensland. This transformation is evident across a range of thresholds, with the shift from distinct clusters to sprawling regional entities occurring at distances beyond 1,480 metres. 

These findings challenge traditional planning paradigms that promote consolidation and compact urban development within these states. Their contiguity, from Brisbane to Melbourne, highlights a significant opportunity for planning and managing metropolitan and regional complexities as critical aspects of the country's future spatial planning. Peripheral growth is becoming as important as urban densification, requiring a governance structure that addresses urban development in concert with broader social and economic policies across the states. This is essential to maintain consistency in urban planning across regions, particularly in the face of significant population growth concentrated mainly in Australia's four largest cities.

Several studies have also recognised this increasing interconnectedness of Australia's eastern crescent and the potential limitations of the current planning system in accommodating these emerging spatial patterns. In particular, \cite{spearritt2009200} and \cite{steele2011governing} both present empirical evidence and similar conclusions regarding the management of the "\textit{200-km city}" encompassing Brisbane, the Gold Coast and the Sunshine Coast --- the same urban cluster identified in this paper with a $d = 1,480m$ and $d = 2,340m$. Building on this, \cite{steele2011governing} argues for a collaborative governance framework that transcends local, regional and national boundaries. This framework would include comprehensive spatial planning and policy integration to address the multiple challenges posed by extensive urbanisation, such as climate change, infrastructure demands, socio-spatial equity and environmental integrity. Such a framework remains critical if Australian planning is to effectively adapt to the evolving urban landscape and manage the complexities of mega-metropolitan regions.

Another key finding of the paper's analysis is the notable functional differentiation between cities that emerges within Australia's urban hierarchy, consistent with \cite{freestone2003functions}'s classification of cities and the empirical findings of \cite{BITRE_2014}. The analysis highlights the significant legacy of agricultural and mining towns in Australia's development. Across the network transition states, distinct towns with a shared history of mining and agricultural services continue to emerge, highlighting the historical importance of these activities in Australia's urban spatial distribution and development. For example, the Goldfields region of Western Australia, several hundred kilometres east of Perth, illustrates the impact of mining on the establishment and peripheral distribution of towns in the state's regional landscape. Similarly, towns such as Karratha, Broome and Mount Isa are notable for their historic role as industrial and mining centres, contributing significantly to their regional economies despite their strong geographical isolation.

However, significant changes in urban centres over the past 20 years have challenged the legacy of agricultural and mining towns in the overall development of the country. While distinct cities with a shared history of mining and agricultural services continue to emerge, highlighting their historical importance in Australia's urban spatial distribution and development, it is worth considering whether \cite{freestone2003functions}'s original labels of the functional differentiation of the country's cities and towns still apply. There may be opportunities to update the characterisation of the many cities and towns that exist in contemporary Austarlia by better understanding the interactions between employment and population networks within these cities and towns. Current prevailing work preferences, particularly for fly-in-fly-out long-distance commuters, have created a dependence of regional centres and townships on transient populations and workforces, raising questions about the sustainability and development of areas reliant on an ever-changing workforce and fluctuating resource economies \cite{hussain2015fly, perry2015fly}.

Australia's urban challenges are characterised by the contrasting approaches of urban containment and extensive regional growth policies. This duality stems from historical continuities and discontinuities, particularly in land use and infrastructure development. While past infrastructure and governance, such as railways and company-owned mining townships, have shaped urban development, they may also limit the flexibility of urban systems to adapt to new economic realities and technological advances. A critical examination of how historical constraints hinder or shape future urban growth is essential to inform more adaptive, forward-looking urban planning strategies. Questions arise as to whether the future planning of autonomous regional cities should include input from larger neighbouring cities. Coordination at the sub-national and regional levels is essential to ensure consistent urban development across states. This approach would integrate regional cities with larger urban centres and promote coherent, sustainable development strategies that address both regional needs and broader metropolitan growth.

We also need to consider their current and future significance and the extent to which they require revitalisation and continued investment. Australia's smaller regional centres and cities are at risk of decline, as highlighted by \cite{maxwell2004mining}, due to the centralising effect of urban centres in Australia's contemporary growth pattern, with larger regional cities providing access to communities and amenities. Their geographical isolation and reliance on specific industries make them vulnerable to economic fluctuations and shifts in industry demand. They serve as critical transition zones, linking regional and urban centres. However, they remain vulnerable to potential declines in the mining and agricultural sectors.Current strategies should prioritise diversified economic approaches and sustainable urban planning to ensure their long-term viability and integration into the wider urban system.Ultimately, the future relevance and survival of these cities are pressing concerns that require further research and strategic planning. Ultimately, the future relevance and survival of these cities remain key concerns that require further research and strategic planning.

Finally, the current trajectory suggests that there are emerging centres that extend across state boundaries, particularly in northern New South Wales, where urban and regional growth are increasingly merging. This requires multi-scalar governance, involving synchronised efforts across all levels of government - local, state and national - as well as contributions from the private and community sectors. Such governance is critical as Australian cities move away from isolated centres and require adaptive infrastructure and planning to strengthen their resilience through sustainable development strategies. These efforts need to be coordinated across all levels of government and supported by the private and community sectors.

\subsection{Limitations and future work}
This study has provided insights into the hierarchical structure of Australian urban centres using percolation analysis. However, it is important to acknowledge that reliance on road network connectivity as the primary data source may not fully capture the complexities of urban development and interactions between regions. Urban systems involve a multitude of factors beyond physical connectivity, including social, economic, and environmental interactions, which may not be adequately represented in this analysis. As such, this paper serves as a foundational starting point, but future research should incorporate additional connectivity measures to capture the multifaceted nature of urban systems. Integrating other forms of connectivity, such as social networks, economic ties, and various transportation modes beyond road networks, could provide a more comprehensive understanding of urban dynamics. 

Further work should also focus on smaller geography analysis to capture localised patterns and nuances in urban development that may be obscured in larger-scale analyses. Future research could address these limitations by breaking down the percolation analysis within individual states to better understand the spatial distribution of towns and the functional differentiation of these towns. This approach would include new classification of urban centres to allow for a more detailed examination of urban structures and take account of their unique geographical and socio-economic development histories. Incorporating this into the percolation clusters would be crucial in capturing the complexity of Australia's urban systems more comprehensively.

\section{Conclusions}
This study utilised percolation theory to explore the hierarchical structures of Australian urban centres, utilising road network connectivity as a proxy. The analysis revealed distinct clusters of urban centres that evolve as connectivity thresholds increase. This methodological approach goes beyond traditional statistical frameworks used in urban studies and provides novel insights into the complex linkages that underpin Australia's diverse urban landscape. 

A notable finding was the interplay between connectivity and separation within Australia's urban system, reflecting the country's unique geography and historical patterns of development. Initially, capital cities and major towns formed distinct clusters, but as connectivity thresholds increased, a pronounced east-west spatial divide emerged, along with the persistent disconnection of historically significant agricultural and mining centres. The study also highlighted the crucial role of regional centres that emerge at intermediate thresholds, acting as vital hubs linking larger urban agglomerations with smaller, isolated communities and facilitating access to goods, services and infrastructure.

The research also observed the transformation of distinct urban clusters across Australia into extensive regional entities, particularly along the densely populated east coast. This evolution challenges traditional planning paradigms focused on consolidation and compact urban development. It highlights the potential emergence of mega-metro regional hybrids, particularly along the eastern crescent spanning cities such as Melbourne, Sydney and Brisbane, where interconnected suburban and urban networks could merge into large regional entities. The pattern of peripheral growth, together with continued urban densification, underlines the need for governance structures that integrate urban development with broader regional and national social and economic policies. Such integrated approaches are essential to ensure coherence and consistency across scales and regions.

The findings underline the growing importance of national coordination in urban development planning. As urban areas potentially extend beyond national borders, a coordinated approach at national level is essential to ensure consistency across regions. Effectively managing the complexities of a potential mega-metropolitan region requires collaborative efforts across local, regional and national boundaries. This is particularly relevant in Australia, where urban growth patterns may not align with existing state boundaries, highlighting the need for collaborative efforts across local, regional and national boundaries.

\section*{Declarations}

The data and materials used in this study are available upon request from the corresponding author.\\

The authors declare that they have no competing interests.\\

No specific funding was received for this research.\\

\textbf{Authors' Contributions:}
\begin{itemize}
    \item \textbf{Dr. Matthew Ng}: Developed the code, devised the analytical framework, conducted the analysis, and contributed to writing.
    \item \textbf{Dr. Zara Shabrina}: Contributed to writing and editing, conducted additional analysis, and reviewed the discussion.
    \item \textbf{Assoc. Prof. Somwrita Sarkar}: Contributed to writing and editing.
    \item \textbf{Prof. Hoon Han}: Contributed to writing and editing.
    \item \textbf{Prof. Christopher Pettit}: Contributed to writing and editing.
\end{itemize}

The authors would like to thank Prof. Elsa Arcaute at the Centre for Advanced Spatial Analysis (CASA), University College London,  for teaching the theory of percolation, demonstrating its implementation, and sharing the code that contributed to this work.

\end{document}